\documentclass[submission, Phys]{SciPost}

\hypersetup{
    colorlinks,
    linkcolor={red!50!black},
    citecolor={blue!50!black},
    urlcolor={blue!80!black}
}

\usepackage[bitstream-charter]{mathdesign}
\DeclareSymbolFont{usualmathcal}{OMS}{cmsy}{m}{n}
\DeclareSymbolFontAlphabet{\mathcal}{usualmathcal}

\begin{document}

\begin{center}{\Large \textbf{
Leptophilic Portals to New Physics at Colliders\\
}}\end{center}

\begin{center}
P. S. Bhupal Dev\textsuperscript{$1\star$}
\end{center}

\begin{center}
$^1$Department of Physics and McDonnell Center for the Space Sciences, \\
Washington University, St.~Louis, MO 63130, USA
\\
* bdev@wustl.edu
\end{center}



\definecolor{palegray}{gray}{0.95}
\begin{center}
\colorbox{palegray}{
  \begin{tabular}{rr}
  \begin{minipage}{0.1\textwidth}
    \includegraphics[width=30mm]{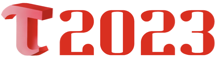}
  \end{minipage}
  &
  \begin{minipage}{0.81\textwidth}
    \begin{center}
    {\it The 17th International Workshop on Tau Lepton Physics}\\
    {\it Louisville, USA, 4-8 December 2023} \\
    \doi{10.21468/SciPostPhysProc.?}\\
    \end{center}
  \end{minipage}
\end{tabular}
}
\end{center}

\section*{Abstract}
{
Observed neutrino oscillations imply that the global lepton flavor symmetry of the Standard Model must be broken. Therefore, searches for lepton flavor violation (LFV) are promising probes of new physics beyond the Standard Model. High-energy colliders provide a powerful tool to study LFV effects, which are complementary to the low-energy charged LFV searches. Here we discuss the possibility of LFV signals at colliders arising from exotic Higgs decays, and from leptophilic scalar and vector portal scenarios~\cite{mytalk}. 
}


\section{Introduction}
\label{sec:intro}
While the Large Hadron Collider (LHC) has consolidated the robustness of the Standard Model (SM), its ultimate purpose of finding beyond-the-SM (BSM) phenomena remains unfulfilled. Instead, stringent bounds, up to several TeV, have been placed on the scale of new physics. This, however, does not necessarily preclude sub-TeV scale BSM particles, e.g.~if they turn out to be {\it leptophilic}, viz.~electrically neutral and only coupling to the SM leptons (and not quarks) at leading order. Here we discuss two such examples of leptophilic scalar and vector resonances at the LHC and future lepton colliders. 

Another interesting aspect of BSM physics
is lepton flavor violation (LFV), which is forbidden in the
SM by an accidental global symmetry $U(1)_B\times U(1)_{L_e}\times U(1)_{L_\mu}\times U(1)_{L_\tau}$. In fact, the observation of neutrino oscillations necessarily implies
LFV at some level. However, the LFV induced just by the light neutrino mass turns out to be negligible~\cite{Petcov:1976ff}. Therefore, any observation of LFV would be a `smoking gun' signal of new physics involving BSM particles~\cite{Davidson:2022jai}. The low-energy charged LFV searches ($\mu\to e\gamma$, $\tau\to 3\ell$, etc) have come a long way in putting some of the most stringent constraints to date. Collider experiments like the LHC provide a powerful complementary probe of LFV at high energies, e.g. via exotic decays of heavy SM particles~\cite{Altmannshofer:2022fvz}, as illustrated below with the SM and BSM Higgs bosons. 

\section{LFV Higgs Decays}
\label{sec:Higgs}
\begin{figure*}[!t]
    \centering
    \includegraphics[width=0.48\textwidth]{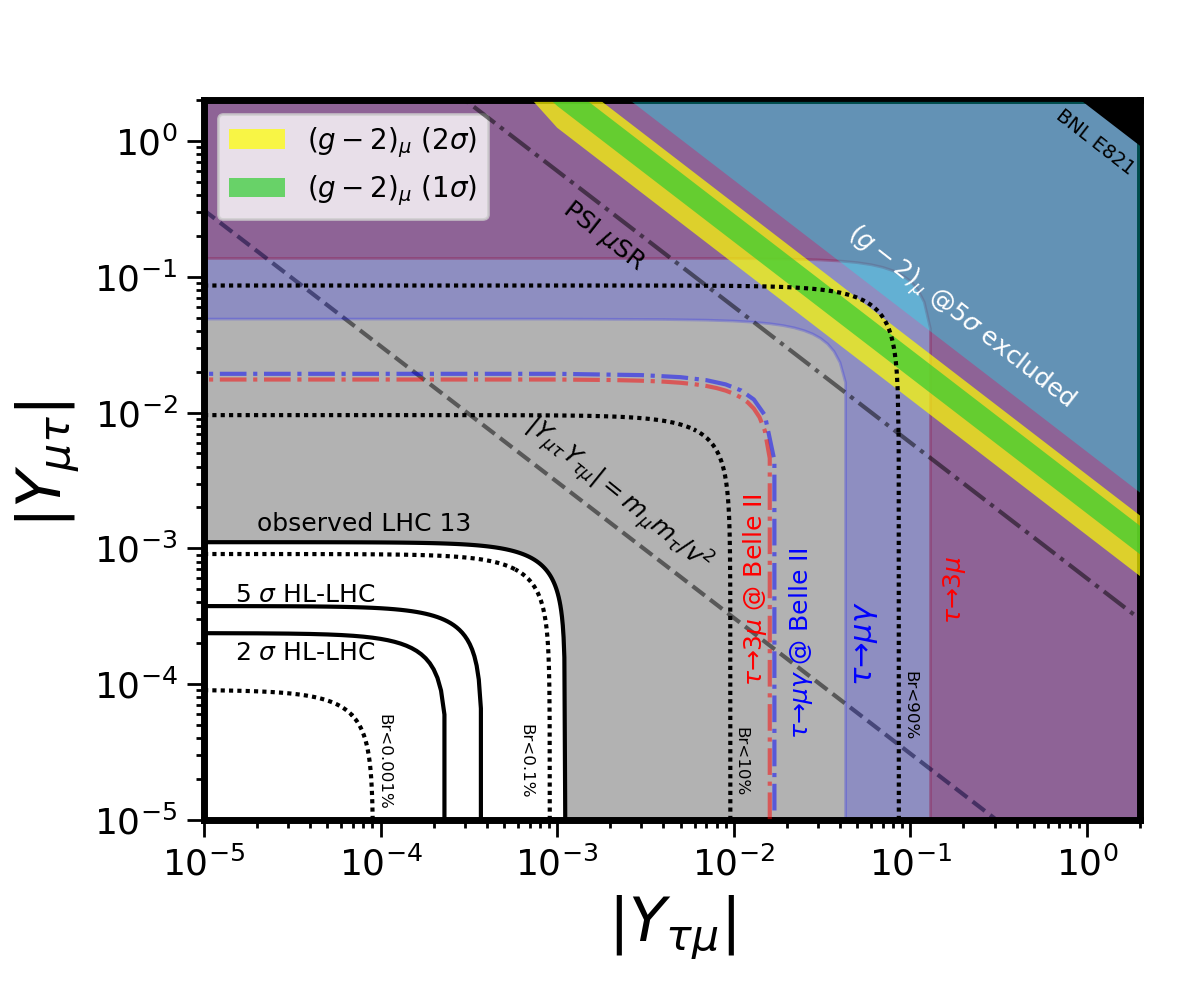}
     \includegraphics[width=0.48\textwidth]{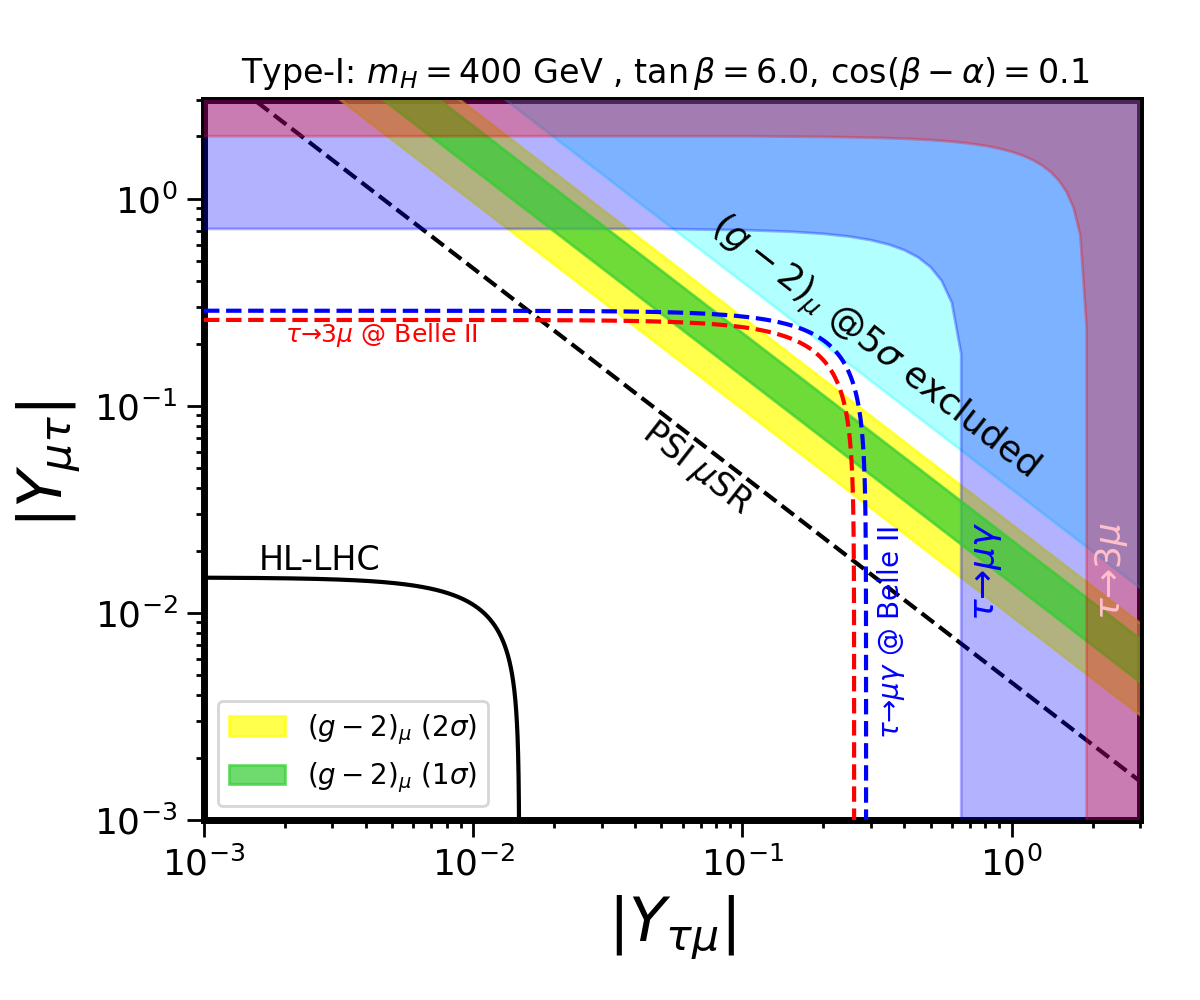}
    \caption{Projected HL-LHC upper limits on the off-diagonal Yukawa couplings $|Y_{\mu\tau}|$ and $|Y_{\tau\mu}|$ from $h/H\to \mu^\pm \tau^\mp$ searches, with $h$ being the SM Higgs boson (left) and $H$ being a heavy Higgs boson in type-I 2HDM (right)~\cite{Barman:2022iwj}. }
    \label{fig:LFV_Sm_Higgs}
\end{figure*}
Within the SM, the Higgs boson couplings to fermions are flavor diagonal. However, in presence of BSM physics, the Yukawa coupling in the mass basis can in general be written as 
\begin{equation}
    Y_{ij} = \frac{m_i}{v} \delta_{ij} + \frac{v^2}{\sqrt{2} \Lambda^2} \lambda_{ij} \, ,
    \label{eq:Yuk}
\end{equation}
where $v$ is the electroweak vacuum expectation value (VEV), $\Lambda$ is the scale of new physics and $\lambda_{ij}$ are the (non-diagonal) coefficients associated with the dimension-6 effective operators that modify the Yukawa interactions~\cite{Herrero-Garcia:2016uab}. This induces LFV Higgs decays $h\to e\mu,e\tau,\mu\tau$ which have been searched for at the LHC~\cite{ATLAS:2019old, CMS:2019pex, CMS:2021rsq, ATLAS:2023mvd, CMS:2023pte}, leading to stringent bounds on the off-diagonal $|Y_{ij}|$, which already surpass those coming from low-energy LFV observables, as shown in Fig.~\ref{fig:LFV_Sm_Higgs} for the $\mu\tau$ case~\cite{Barman:2022iwj}. The HL-LHC will further improve these bounds by a factor of few.

\section{Leptophilic Neutral Scalars}
Although no evidence for LFV decays of the 125~GeV Higgs boson was found at the LHC, CMS has reported an intriguing $3.8\sigma$ local ($2.8\sigma$ global) excess in the resonant $e\mu$ search around 146~GeV, with a preferred cross section of $\sigma(pp\to H \to e\mu)=3.89^{+1.25}_{-1.13}$~fb~\cite{CMS:2023pte}. Although a similar excess was not found in the ATLAS search~\cite{ATLAS:2019old}, which disfavors the CMS excess at $1\sigma$~\cite{Leney}, it should be noted that unlike the CMS analysis, ATLAS did not perform a dedicated, boosted decision tree (BDT)-optimized resonance search, and did not interpret the results for masses different than 125 GeV. The nature of the CMS excess is therefore unclear and will have to be clarified with more data collected in the next run.

\begin{figure*}[!t]
    \centering
    \includegraphics[width=0.45\textwidth]{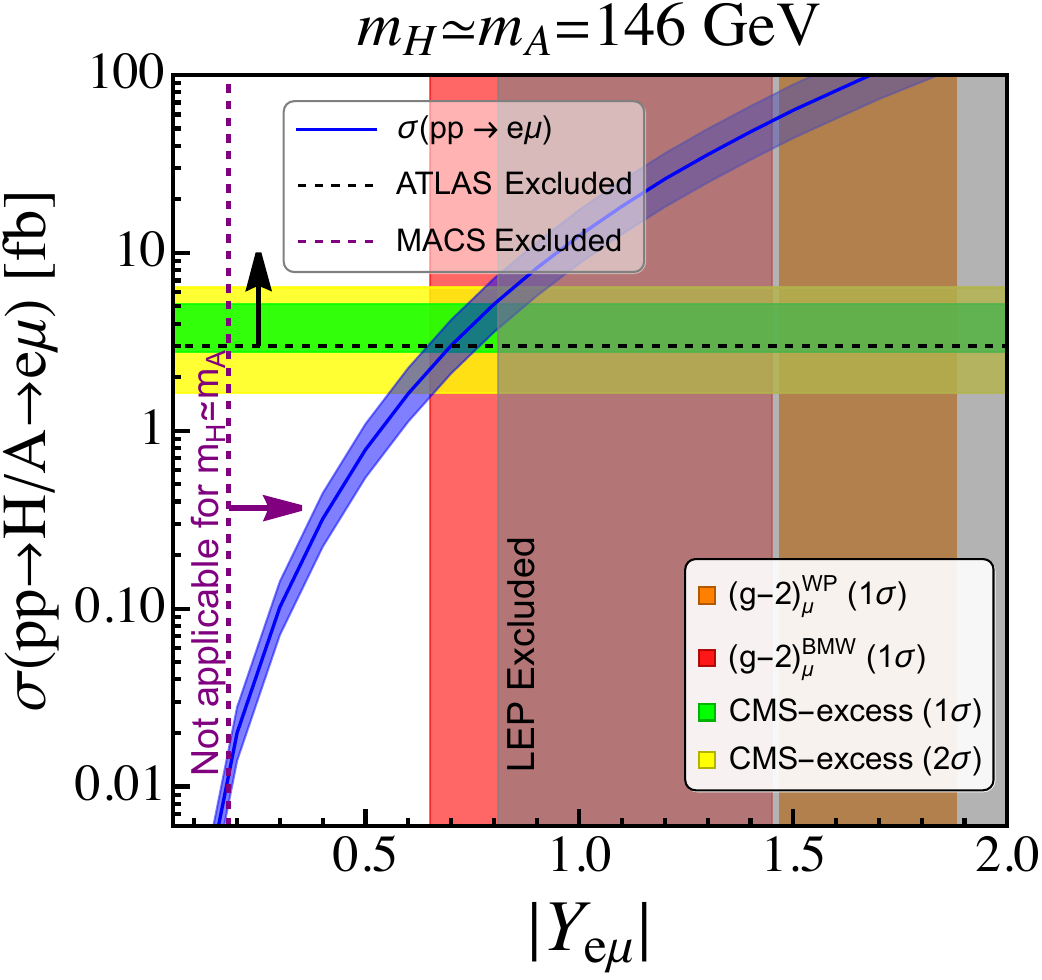} 
    \includegraphics[width=0.48\textwidth]{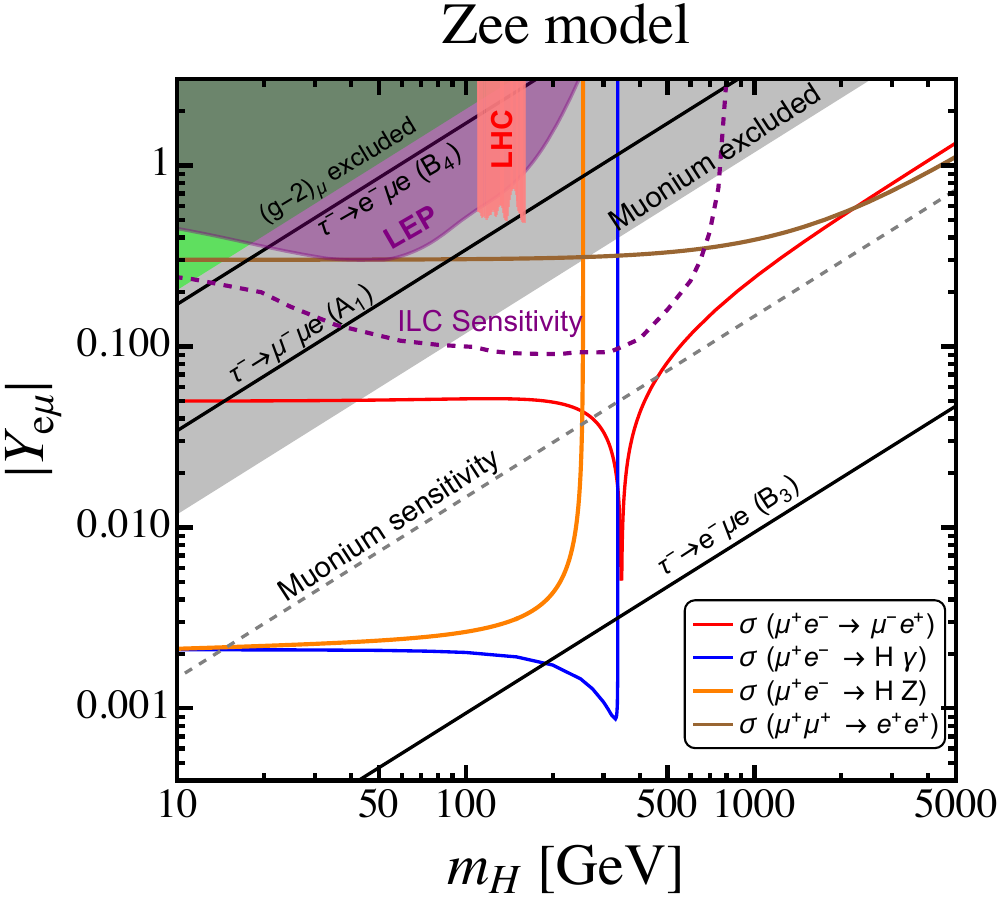}
    \caption{{\it Left:} Fitting the CMS $e\mu$ excess in a leptophilic 2HDM~\cite{Afik:2023vyl}. {\it Right:} Future lepton collider sensitivities to the Zee model parameter space~\cite{Dev:2023nha}.
    } 
    \label{fig:p2}
\end{figure*}

Taking the CMS $e\mu$ excess at face value, we provide the {\it simplest} possible interpretation in terms of a leptophilic neutral scalar within a two-Higgs-doublet model (2HDM)~\cite{Afik:2023vyl}. The relevant Yukawa interactions of the second Higgs doublet are given by 
\begin{equation}
        - {\cal L}_Y \supset Y_{\alpha\beta} \bar{L}_\alpha H_2 \ell_{\beta,R} +  {\rm H.c.} \, , \quad {\rm where}~H_{2}=\left(\begin{array}{c}
H^{+} \\
\frac{1}{\sqrt{2}}\left(H+i A\right) 
\end{array}\right) \, .
        \label{eq:Yuk2HDM}
    \end{equation}
    The resonant $e\mu$ production at the LHC ($pp\to H/A\to e\mu$) is then attributed to the lepton parton distribution function of the proton~\cite{Bertone:2015lqa, Buonocore:2020nai, Dreiner:2021ext}. This is different from other scalar interpretations of the CMS excess~\cite{Primulando:2023ugc, Koivunen:2023led, Bhattacharya:2023lmu}, which used quark couplings to enhance the production cross section via gluon fusion. As shown in Fig.~\ref{fig:p2} (left), our leptophilic scenario can explain the CMS excess with a Yukawa coupling $Y_{e\mu}\sim 0.55$--0.81, while being consistent with all existing constraints. The killer constraint from muonium-antimuonium oscillation~\cite{Willmann:1998gd} can be evaded by making the CP-even and CP-odd neutral scalars quasi-degenerate in mass. Interestingly, the same $Y_{e\mu}$ coupling is also consistent with the muon $(g-2)$ anomaly~\cite{Muong-2:2023cdq}, if we take the BMW value~\cite{Borsanyi:2020mff} for the SM prediction. Moreover, by introducing a mass splitting between the charged and neutral components of the second Higgs doublet in this leptophilic 2HDM, we can also explain the CDF $W$-mass anomaly~\cite{CDF:2022hxs}.

Irrespective of the future status of these anomalies, the prospects of probing a leptophilic light Higgs sector at the energy and intensity frontiers is a worthwhile study in its own right. One of the reasons is that leptophilic scalars arise naturally in radiative neutrino mass models~\cite{Cai:2017jrq}. Taking the Zee model~\cite{Zee:1980ai, Babu:2019mfe} as a prototypical example, we show in Fig.~\ref{fig:p2} (right panel) some future lepton collider sensitivities~\cite{Dev:2023nha}, along with the existing constraints from LEP and LHC, as well as from low-energy LFV searches. Note that the tau LFV constraints depend on the Yukawa texture and can be rendered much weaker than the collider limits.      

\section{Leptophilic Vector Bosons}
There are good symmetry reasons that can motivate leptophilic neutral gauge bosons ($Z'$). First of all, the presence of additional Abelian symmetries like $U(1)$ often arise as low-energy remnants of Grand Unified Theories, string compactifications, or extra dimensional models at a high scale~\cite{Langacker:2008yv}, which always come with the corresponding $Z'$ bosons after the $U(1)$-symmetry breaking. As for its leptophilic nature, a simple anomaly-free local $U(1)$ gauge extension of the SM that realizes it is $U(1)_{L_\alpha-L_\beta}$ (where $\alpha,\beta=e,\mu,\tau$ with $\alpha\neq \beta$)~\cite{Foot:1990mn, He:1991qd, Foot:1994vd}. 
 The relevant terms in the effective  Lagrangian are given by 
\begin{equation}
-\mathcal{L}_{L_\alpha-L_\beta} \supset g' Z'_\mu (
\bar{L}_{\alpha}\gamma^\mu L_\alpha-\bar{L}_\beta\gamma^\mu L_\beta+\bar{\ell}_{R\alpha}\gamma^\mu \ell_{R\alpha}
-\bar{\ell}_{R\beta}\gamma^\mu e_{\ell\beta})
+\frac{1}{2}M_{Z'}^2Z'_\mu Z'^{\mu},
\end{equation} 
where the $Z'$ mass is given by $M_{Z'}=2g'v_\Phi$, with $v_\Phi$ being the VEV of the extra scalar field $\Phi$ which is singlet under the SM gauge group but has a charge 2 under $U(1)_{L_\alpha-L_\beta}$.

\begin{figure}[!t]
\centering
\includegraphics[width=0.32\textwidth]{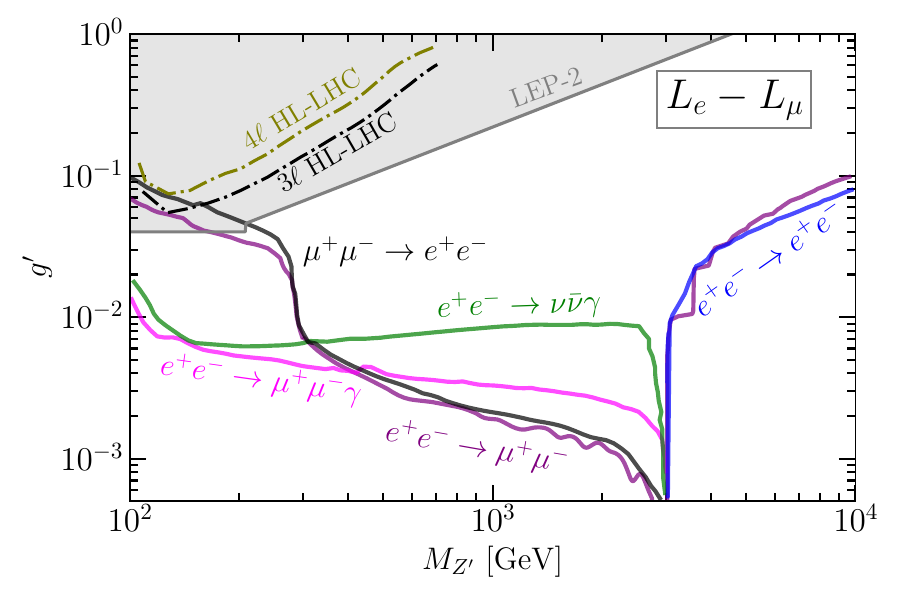}
\includegraphics[width=0.32\textwidth]{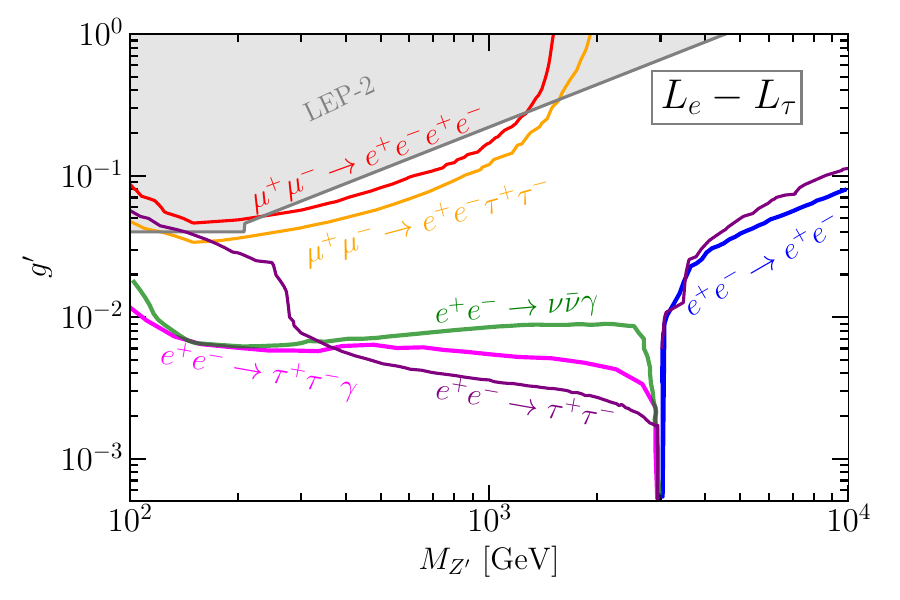}
\includegraphics[width=0.32\textwidth]{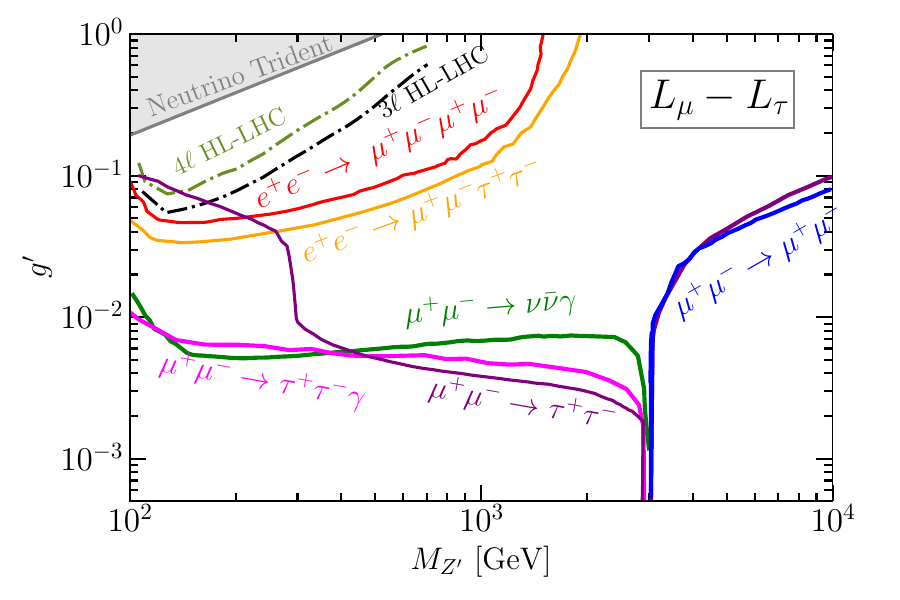}
\caption{The $95\%$ CL sensitivity for the $U(1)_{L_\alpha-L_\beta}$ models at $\sqrt{s} = 3~{\rm TeV}$ electron and muon colliders~\cite{Dasgupta:2023zrh}. 
}
\label{fig:sensi}
\end{figure}

Since the quark couplings of the $Z'$ in this model are induced only at the loop level, the most stringent dijet constraints on heavy $Z'$ coming from Tevatron and LHC~\cite{Dobrescu:2021vak} are not applicable, thus opening up a large chunk of parameter space in the $Z'$ mass-coupling plane. Future lepton colliders provide an ideal environment for probing this unexplored leptophilic $Z'$ parameter space at the electroweak scale and above. This is shown in Fig.~\ref{fig:sensi}, taking $\sqrt s=3$ TeV future $e^+e^-$ and muon colliders as examples~\cite{Dasgupta:2023zrh}. The HL-LHC sensitivities in the $3\ell/4\ell$ channels~\cite{delAguila:2014soa} are also shown for comparison, along with the existing constraints (grey shaded regions). Note that in the allowed parameter space shown here, the leptophilic $Z'$ can also serve as a portal to the dark sector, with very interesting phenomenology of its own~\cite{Blanco:2019hah}. 

Another interesting and complementary aspect of the leptophilic $U(1)$ models pertains to the cosmological phase transition of the $U(1)$-symmetry-breaking scalar field $\Phi$. If the symmetry is classically conformal, the tree-level potential is flat due to scale-invariance, and thermal corrections can easily dominate to make the phase transition strongly first order, leading to a potentially observable stochastic gravitational wave (GW) signal. In fact, the current GW data from advanced LIGO/VIRGO~\cite{KAGRA:2021kbb} already excludes a portion of the $U(1)_{L_\alpha-L_\beta}$ model parameter space at high $M_{Z'}$ values not accessible to colliders, whereas the next-generation GW experiments in the mHz-kHz regime, such as $\mu$ARES, LISA, DECIGO, BBO, ET, and CE will further extend the sensitivity reach to complement the collider reach presented in Fig.~\ref{fig:sensi}~\cite{Dasgupta:2023zrh}.

\section{Conclusion}
Any observation of LFV signal involving charged leptons would be a `smoking gun' for BSM physics. While the low-energy charged LFV searches must continue, the high-energy LFV signals at colliders provide a powerful complementary probe of a wide variety of BSM scenarios. Here we discussed a few such examples of LFV in exotic Higgs decays and from leptophilic scalar or vector bosons. In each case, we showed that LFV searches at future colliders, from HL-LHC to a future electron/muon collider, can provide new insights into BSM physics. We should keep a close eye on the intriguing hint of an LFV signal emerging from the LHC data.  

\section*{Acknowledgments}
The work of PSBD is supported by the U.S.~Department of Energy under Grant No.~DE-SC0017987.

\bibliography{TAU2023_Dev.bib}

\nolinenumbers

\end{document}